\documentclass[fleqn,usenatbib]{mnras}

\usepackage{newtxtext,newtxmath}

\usepackage[T1]{fontenc}

\DeclareRobustCommand{\VAN}[3]{#2}
\let\VANthebibliography\thebibliography
\def\thebibliography{\DeclareRobustCommand{\VAN}[3]{##3}\VANthebibliography}

\usepackage{graphicx}	
\usepackage{amsmath}	
\usepackage{amssymb}	
\usepackage{dcolumn}
 
\title{Semi-analytical frameworks for subhalos from the smallest to the largest scale}

\author[N. Hiroshima et al.]{Nagisa Hiroshima, $^{1,2}$
Shin'ichiro Ando,$^{3,4}$
and Tomoaki Ishiyama$^5$ \\ 
$^{1}$Department of Physics, University of Toyama, 3190 Gofuku, Toyama 930-8555, Japan\\
$^{2}$RIKEN iTHEMS, Wako, Saitama 351-0198, Japan\\
$^{3}$GRAPPA Institute, Institute of Physics, University of Amsterdam, 1098 XH Amsterdam, The Netherlands\\
$^{4}$Kavli Institute for the Physics and Mathematics of the Universe (WPI), University of Tokyo, Kashiwa 277-8583, Japan\\
$^{5}$Institute of Management and Information Technologies, Chiba University, Chiba 263-8522, Japan }

\begin{document}
\label{firstpage}
\pagerange{\pageref{firstpage}--\pageref{lastpage}}
\maketitle

\begin{abstract}
Substructures of dark matter halo, called subhalos, provide important clues to understand the nature of dark matter. We construct a useful model to describe the properties of subhalo mass functions based on the well-known analytical prescriptions, the extended Press-Schechter theory. The unevolved subhalo mass functions at arbitrary mass scales become describable without introducing free parameters. The different host halo evolution histories are directly recast to their subhalo mass functions. As applications, we quantify the effects from (i) the Poisson fluctuation, (ii) the host mass scatter, and the (iii) different tidal evolution models on observables in the current Universe with this scheme. The Poisson fluctuation dominates in the number count of the mass ratio to the host of $\sim {\cal O}(10^{-2})$, where the intrinsic scatter is smaller by a factor of a few. The host-mass scatter around its mean does not affect the subhalo mass function. Different models of the tidal evolution predict a factor of $\sim2$ difference in numbers of subhalos with $\la {\cal O}(10^{-5})$, while the dependence of the Poisson fluctuation on the tidal evolution models is subtle. The scheme provides a new tool for investigating the smallest-scale structures of our Universe which are to be observed in near future experiments.
\end{abstract}

\section{Introduction}
\label{s:intro}
The nature of dark matter (DM) is still elusive, while the evidence for its existence is strong (e.g. rotation curves of galaxies~\citep{vanAlbada:1984js}, bullet clusters~\citep{Clowe:2003tk}, etc.). One of the strongest, and the original motivations for DM comes from the structure formation history of our Universe~\citep{Peebles:1982ff,Planck:2018vyg}. New particles beyond the Standard Model as well as non-particle solutions, such as primordial black holes, which source the gravitational potential for baryonic structures to evolve are required to explain the current structures. Various possibilities are considered for solving this long-standing problem (see \citet{Bertone:2004pz,
Bertone:2018krk,Carr:2020gox} for reviews).

The hierarchical structure of halos that form through the gravitational collapse of matter overdensities is a key to access the nature of DM. The structure formation proceeds through mergers, accretion, and tidal stripping. Its statistical property reflects the nature of particle DM. For example, the minimum halo mass of the halo ranges from ${\cal O}(10^{-12})$--${\cal O}(10^{-3})M_\odot$ depending on parameters of the weakly interacting massive particles~\citep{Diamanti:2015kma,Bringmann:2009vf,Hofmann:2001bi,Green:2003un,Loeb:2005pm,Profumo:2006bv}. Introduction of the self-interaction in DM sector or velocity dispersion could erase the small-scale structures~\citep{Spergel:1999mh,Buckley:2014hja,Nadler:2020ulu,Bond:1980ha,Hogan:2000bv}. The importance of the clustering of the small-scale halos in its host halo appears in another context of the indirect searches of DM. The annihilation signals of DM should be enhanced by a factor of several due to the clustering effects, which are known as the boost factor~\citep{Sanchez-Conde:2013yxa,Ishiyama:2014uoa,Moline:2016pbm,Hiroshima:2018kfv,Ishiyama:2019hmh}. The determination of the slope of the subhalo mass function as well as the minimum halo mass is crucial in its estimation (see \citet{Ando:2019xlm} for a recent review). 

Numerical~\citep{Gao:2004au,Diemand:2007qr,Giocoli:2007uv,Ishiyama:2020vao,Moline:2021rza}, semi-analytical~\citep{Penarrubia:2004et}, and analytical~\citep{Giocoli:2007gf,vandenBosch:2004zs,Yang:2011rf} studies have investigated the properties of subhalos such as their mass functions. Numerical simulations are advantageous in treating all effects determining the properties, while they are limited by numerical resolutions and computational costs. Semi-analytical models cure such difficulties by approximating several parts of the physical processes. The merger tree based on the extended Press-Schechter (EPS) theory~\citep{Bond:1990iw,Lacey:1993iv}, which is a theoretically-motivated extension of the original Press-Schechter model~\citep{Press:1973iz}, takes an essential role in the semi-analytical approaches. 

Previous studies have developed schemes to reconstruct the so-called unevolved mass function, which is the mass function corresponding to that before the tidal evolution, in semi-analytical ways. Based on the EPS prescription, \citet{Zhao:2008wd} has derived a fitting formula to follow the main branch of the merger tree calibrated against the numerical simulation. The main branch of a halo is obtained by repeatedly connecting the most massive progenitor of the most massive progenitor. The work by \citet{Zhao:2008wd} is extended in \citet{Yang:2011rf} to include the scatter around the main branch of the tree, showing that the unevolved subhalo mass function is well-accessible in this scheme. On the other hand,  \citet{Correa:2014xma,Correa:2015dva} have derived analytical formulae describing the evolution of the main branch halos in a way that does not require the fitting to the simulation results. It is achieved by examining the growth rate of the initial density perturbations. 

In order to access the astrophysical observables such as the number counts of the satellite galaxies, those models must be implemented with the prescription for tidal evolution after the subhalo accretion, as pointed out in \citet{Giocoli:2007gf}. \citet{vandenBosch:2004zs} has proposed to adopt the orbit-averaged mass-loss rate to treat the tidal evolution in a semi-analytical way. \citet{Jiang:2014nsa} calculate the mass-loss rate by segmenting the time steps for treating the host evolution. The free parameters of the mass-loss model are tuned to reproduce the evolved subhalo mass function in numerical simulations in their succeeding works~\citep{Jiang:2013kza,Jiang:2014nsa}. The normalization of the subhalo mass function before the tidal evolution had to be fitted using numerical results as those in previous works~\citep{Giocoli:2007uv,Li:2009ky}.

In this work, we propose an EPS-based analytical scheme to access the statistics of subhalos making full use of its advantage. The construction of the unevolved subhalo mass function in this scheme requires no parameters to be fitted. The evolved subhalo mass function is obtained by implementing the orbit-averaged tidal evolution model to the unevolved mass functions, which is a prescription proposed in \citet{vandenBosch:2004zs,Jiang:2013kza,Jiang:2014nsa}. 
Unlike the earlier work that focused on constructing full merger trees including second and third order branches, we only follow the main branch of the merger tree. 
This approach enables us to construct the subhalo mass functions without being limited by the numerical resolution. 
We can fully adopt the analytical formula of the subhalo accretion rates combined with the host evolution, both of which are described by the EPS theory. 
The downside of this approach is incapability of tracing higher order subhalos: i.e., subhalos in a parent subhalo. However, our interest in this work is the properties of subhalos at the first order. Those at higher orders are still very difficult to resolve with numerical simulations or to identify observationally.
As we explain below, we evaluate the subhalo statistics using 500 of host realizations, which reasonably treats the halo-to-halo scatter. In our calculation, the computational cost is highly reduced and  an unlimited range of the subhalo-to-halo mass ratio can be covered simultaneously. We quantify the numerical or observational effects of host-mass scatter and the Poisson fluctuation, which we need to manage in extracting the indication about the intrinsic properties of DM as well as the dependence of the observables on the tidal-evolution model based on this work.

This work improves the host-subhalo connection from our previous works~\citep{Hiroshima:2018kfv,Ando:2019xlm} which provides an access to the subhalo observables adopting semi-analytical method. 
In the following, we directly connect the different host evolution histories to those of subhalo mass functions. This reduces the needs to rely on fitting formulae calibrated against numerical simulations and the accessibility to the statistical quantities is enhanced. Also, the scheme paves the way for halo mass functions in much general cosmological setups.

The structure of this article is as follows. Sec.~\ref{s:sim} is devoted for the simulation we have performed to compare with the analytical calculations. After summarizing the host evolution in the EPS framework in Sec.~\ref{s:host}, we show the prediction about the subhalo mass function in Sec.~\ref{s:shmf}. The scatter of the subhalo mass function due to the extrinsic effects is investigated in Sec.~\ref{s:application}. The dependence on the tidal-evolution model is also discussed in this section. We finally conclude in Sec.~\ref{s:summary}. Throughout the paper, we use ``$\ln$" and ``$\log$" as notations for natural and 10-based logarithmic quantity. We assume cosmological parameters of \citet{Planck:2015fie} (Table 4, “TT+lowP+lensing”) unless mentioned specifically.

\section{Numerical simulation}
\label{s:sim}
In this section, we explain our numerical simulations used for deriving the host evolution and subhalo mass function to compare with the EPS-based calculations. Numerical results provided in this section appear in Secs.~\ref{s:host}, \ref{s:shmf}, and \ref{s:application}.

We use a high-resolution cosmological $N$-body simulation, Shin-Uchuu~\citep{Ishiyama:2020vao}. The Shin-Uchuu simulation consists of $6400^3$ dark matter particles with a mass of $8.97 \times 10^5 \, h^{-1} M_{\odot}$ in a comoving cubic box with a side length of 
140$ \, h^{-1}  \rm Mpc$. The gravitational softening length is 
0.4$ \, h^{-1}  \rm kpc$. The cosmological parameters are $\Omega_{0}=0.3089$, $\Omega_{\rm b}=0.0486$, $\lambda_{0}=0.6911$, $h=0.6774$, $n_{\rm s}=0.9667$, and $\sigma_{8}=0.8159$,
which are consistent with observational results of cosmic microwave background
by the Planck satellite~\citep{Planck:2018vyg}. 

\textsc{rockstar} halo/subhalo finder~\citep{Behroozi:2011ju} was used to find gravitational bound halos and subhalos. Then \textsc{consistent trees} merger tree code~\citep{Behroozi:2011js} was used to construct their merger trees. Those data are available on Skies \& Universes site. 
Further details of the Shin-Uchuu simulation and its basic halo and subhalo properties are presented in the literature~\citep{Ishiyama:2020vao,Moline:2021rza}.

\section{Host halo evolution}
\label{s:host}
Our analytical calculation of the host evolution is based on the prescription in \citet{Yang:2011rf}. We summarize the scheme in this section. Focusing on the Milky-Way-like host at $z=0$, we follow the evolution path of the halo backward in time up to $z=10$. 
Concretely, we follow the steps below.
\begin{enumerate}
\item We set the host mass at $z=0$ and prepare $N_{\rm host}$ of host halos. We use $N_{\rm host}=500$ halo realizations for each single mass host in our calculations. We set our fiducial host mass as $M_{\rm host}(z=0)=1.30\times10^{12}M_\odot$, adopting the convention of the $M_{200}$ that is the mass enclosed in a radius $r_{200}$ within which the average density is 200 times the critical density. This convention matches those in the literature we refer to in our calculations.
\item We prepare the steps of the redshift to sample the host mass. We take a constant interval in the $\ln(1+z)$ space taking $N_z=800$ points from $z = 0$ to $z=10$. In the following, the step size is denoted with $\Delta z$.
\item The distribution of the accreting mass is determined using the EPS formula by specifying the redshift and the host mass. Following \citet{Yang:2011rf}, the number distribution of the subhalos that accreted in the redshift interval between $z+\Delta z$ and $z$ is given as
\begin{eqnarray}
\label{yang5}
F(m)=&&f(s(m),\delta(z+\Delta z)|S(M),\delta(z))\nonumber \\ 
&&\times G\left(\frac{\sqrt{S(M)}}{\sqrt{s(m)}},\frac{\delta(z+\Delta z)}{\sqrt{s(m)}}\right)\left|\frac{ds}{dm}(m)\right|\left(\frac{M}{m}\right)
\end{eqnarray}
where
\begin{eqnarray}
    &&f(s(m),\delta(z+\Delta z)|S(M),\delta(z)) \nonumber \\
    &&=\frac{1}{\sqrt{2\pi}}\frac{\delta(z+\Delta z)-\delta(z)}{\left[s(m)-S(M)\right]^{3/2}}\exp\left[-\frac{\left(\delta(z+\Delta z)-\delta(z)\right)^2}{2(s(m)-S(M))}\right].
\end{eqnarray}
Eq.~\ref{yang5} gives the number of the subhalo in the mass range of $[m,\ m+dm]$ that have accreted in the redshift interval of $[z+\Delta z,\ z]$, i.e. the time interval corresponding to the host evolution from $M(z+\Delta z)$ to $M(z)$. The factor $M/m$ appears in the last part of the equation to obtain the number fraction of mass $m$ from its mass fraction. 
Each quantity in  Eq.~\ref{yang5} is:
\begin{itemize}
\item $\delta(z)=1.686/D(z)$: the linear growth factor.
\item $S(M)$ and $s(m)$: the variance of the linear density field, $\sigma^2(M)$ and $\sigma^2(m)$ at $z=0$. Each of the mass scale $M$ and $m$ corresponds to that of the host and accreting halo at that redshift, respectively.
\item $G(x,y)=G_0 x^{\gamma_1} y^{\gamma_2}$: a correction function for the EPS formula tuned against the numerical simulations.
\end{itemize}

We adopt the parametrized form of the $D(z)$ and $\sigma(m)$ provided in \citet{Ludlow:2016ifl}. 
The parameter values of the correction function $G$ are set to $(G_0,\ \gamma_1,\ \gamma_2)=(0.57,\ 0.38,\ -0.01)$, which we have taken from \citet{Parkinson:2007yh} and \citet{Cole:2007yc}. Those works have tuned the parameters with the Millennium $N$-body simulation~\citep{Springel:2005nw} which covers the halo mass in the range of $[\sim10^{10}h^{-1}M_\odot,\ \sim10^{16}h^{-1}M_\odot]$ and redshift up to $z\sim10$, and suitable for generating the merger tree~\citep{Jiang:2013kza} . 

We regard the most massive subhalo in the redshift interval $[z,z+\Delta z]$ as the progenitor of the host of $M(z)$ at $z$. When the accreting halo mass $m$ of Eq.~\ref{yang5} satisfies $m>M(z)/2$, the hierarchy of halo assures that the halo to be the unique progenitor. We determine the evolution history by following the progenitor adopting the inverse transform method.
\item Prepare subhalos in the mass range of $0.5M(z)\leq m <M(z)$ taking $10^5$ points in linear space.
\item Construct the cumulative distribution function by integrating Eq.~\ref{yang5} in the above mass range, then normalize it. 
\item Generate a random number following the uniform distribution in the interval of $[0,1]$, which are assigned to a point in the normalized cumulative distribution function in the above calculation step. We identify the progenitor at $z+\Delta z$ as that at the corresponding mass in the normalized cumulative mass function.
\item Repeat the same procedure succeedingly to redshift $z=10$. 
\end{enumerate}

\begin{figure}
\begin{center}
\includegraphics[bb=7 10 428 307,scale=0.5]{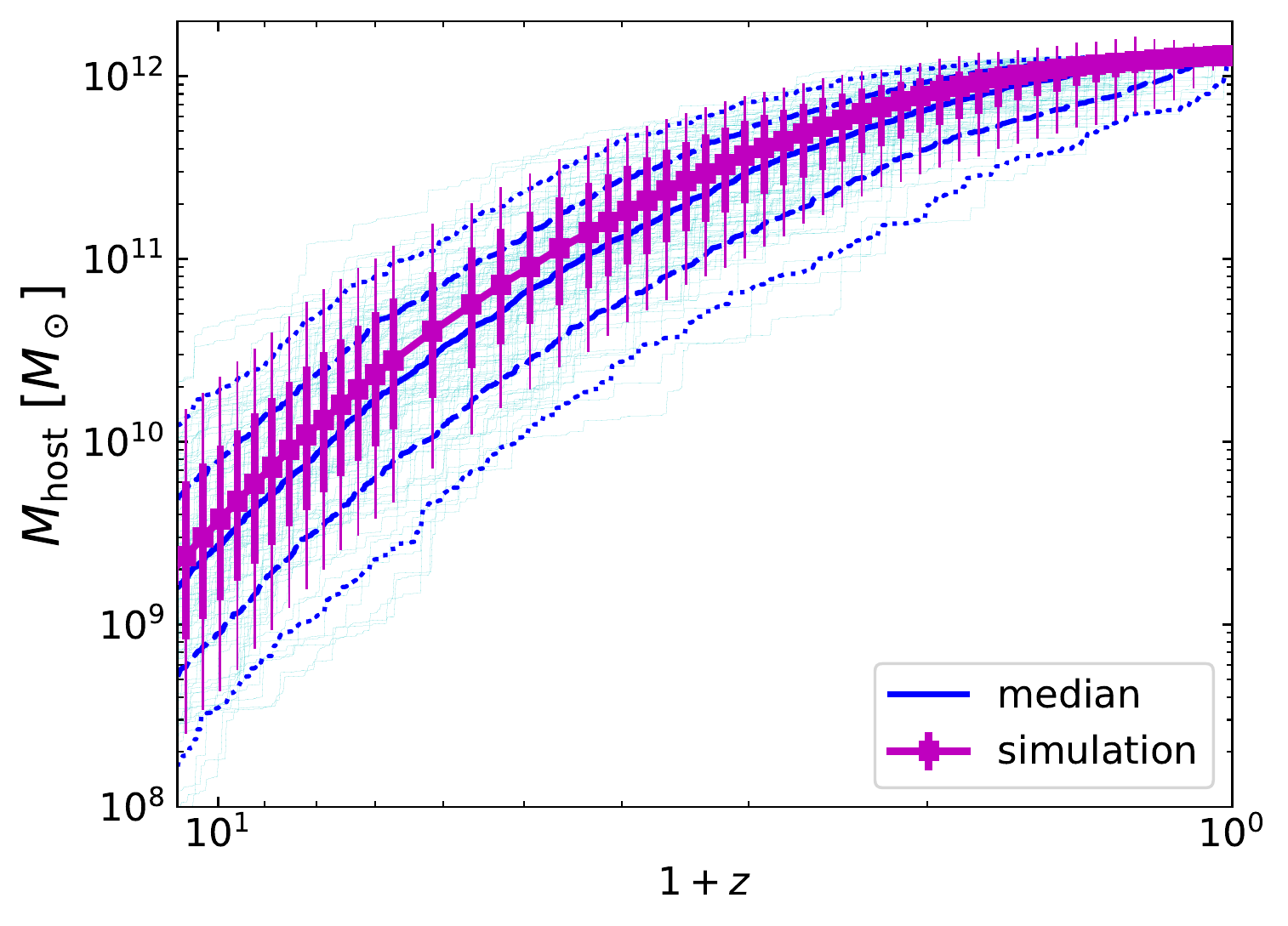}
\caption{Evolution history determined in the EPS formalism for hosts of $M=1.30\times10^{12}M_\odot$ at $z=0$. The blue solid line is the median. Dashed (dotted) lines correspond to 16\% and 84\% (2.5\% and 97.5\%) quantiles. Each realization of the host is shown in thin (cyan) lines. We only plots 100 of the total 500 host realizations in order for the visibility. Numerical results are shown in square points, of which thick (thin) error bars represent the quantiles same as dashed (dotted) lines for EPS-based analytical calculations.  The analytical calculation slightly underestimates the median of the history while it agrees with the simulation within the range of the error bar. }
\label{f:hostevolution}
\end{center}
\end{figure}

In Fig.~\ref{f:hostevolution}, we show the resulting evolution of our fiducial hosts. The solid line is the median of the $N_{\rm host}=500$ realizations and thin lines are the histories of 100 host realizations. We also show the 16\% (2.5\%) and 84\% (97.5\%) quantiles with dashed (dotted) lines. Square points with error bars are the results obtained in numerical simulations which we have explained in Sec.~\ref{s:sim}. As it is seen in the figure, analytical and numerical calculations agree within the range of the error bar in both the median and scatter while the median in the analytical model is slightly smaller.  The host evolution history also agrees with the fitting function in \citet{Correa:2014xma} in the regime where the function is tuned against simulations. In this regime, the functional form is similar to that predicted in \citet{Wechsler:2001cs} and similar results are obtained also in previous works~( \citet{Bosch:2001ej,Zhao:2008wd,Giocoli:2011hz}.) 
All the $N_{\rm host}=500$ realization of the host evolution history is used to derive the subhalo mass function in the next section. 

\section{Subhalo mass function}
\label{s:shmf}
In this section, we explain our self-consistent framework to derive subhalo mass functions reflecting the different host evolution histories. We start from constructing unevolved mass functions based on the host evolution history (see Sec.~\ref{s:host}) in Sec.~\ref{ss:unev}. Assuming that the increment of the host mass is solely determined by accretions of smaller halos, the normalization of the unevolved subhalo mass function at each redshift step is determined without introducing any free parameters. The unevolved mass function at $z=0$ is then obtained by summing up contributions from accreting halos in all redshift steps. The evolved subhalo mass function, which directly relates to the observables in the current Universe, is calculated by combining the unevolved mass function with the tidal-evolution model in \citet{Hiroshima:2018kfv}. Since we hold the mass of the host in each redshift step and realization, the host-evolution dependent mass-loss becomes calculable, as it is explained in Sec.~\ref{ss:ev}. 

\subsection{Unevolved mass function}
\label{ss:unev}
\begin{enumerate}
\item Set a single value for the host halo $M_{\rm host}$ at $z=0$. 
\item Generate the subhalo sample of which mass ranges of $[m_{\rm min},\ m_{\rm max}]=[10^{-18},\ 1]\times M_{\rm host}(z=0)$ for each step of the redshift and host realization. The minimum mass of the sample subhalo can be arbitrarily chosen and we set the value so that to match the prediction in neutralino DM scenarios~\citep{Green:2003un}. We take $N_{\rm sh}=300$ sampling points for the accreting subhalos. 
The interval of the accreting mass sampling is constant in the $\ln(m_{\rm 200})$ space. 
The virial-quantities are used for calculating the evolved mass function as will explain in Sec.~\ref{ss:ev}.
\item Determine the total accreted mass during the interval of $[z+\Delta z,z]$ as $M_{\rm acc}(z)=M(z)-M(z+\Delta z)$ by directly using the host evolution data.
\item Count the number of subhalo in a similar manner with that explained in Sec.~\ref{s:host}. Taking the $\delta(z)$ and $s(m)=\sigma^2(m,z=0)$ as the variables for the redshift and mass, the distribution function is
\begin{eqnarray}
\label{eq:dndm}
\left.\frac{dN}{dm}\right|_{z} =&&f(s(m),\delta(z+\Delta z)|S(M),\delta(z)) \nonumber \\
&&\times G\left(\frac{\sqrt{S(M)}}{\sqrt{s(m)}},\frac{\delta(z+\Delta z)}{\sqrt{s(m)}}\right)\frac{1}{m}\left|\frac{ds}{dm}\right|\frac{dM}{(1+z)},
\end{eqnarray}
where $dM=M(z)-M(z+\Delta z)=M_{\rm acc}(z)$.
Eq.~\ref{eq:dndm} gives the distribution of the accreting subhalo at the redshift step $z$. 
As it is seen from the functional form of Eq.~\ref{eq:dndm}, the distribution function diverges at $m\sim M(z)$. We discard the regime where the distribution function increases with $m$, regarding that such a regime is not physical.\footnote{Our treatment is close to that proposed as the Model III in \citet{Yang:2011rf} in practice, while we do not need to introduce extensional parameters.}
\item Normalize the distribution $dN/dm$ by applying the condition

\begin{equation}
M_{\rm acc}(z) = \sum m(z)dm\left.\frac{dN}{dm}\right|_{z}.
\label{massfcn_norm}
\end{equation}
The sum is taken for subhalos remaining after the cut in the previous step. 
\item Integrate the mass function over the redshift interval of $z=[10,0]$.
\end{enumerate}
The above procedure of the calculation enables us to obtain the unevolved mass functions reflecting the different history of host evolutions. There are no needs for tuning parameters in the above calculation procedure.

\begin{figure*}[t]
\begin{tabular}{cc}
\begin{minipage}{0.45\hsize}
    \includegraphics[bb=7 7 428 311,scale=0.5]{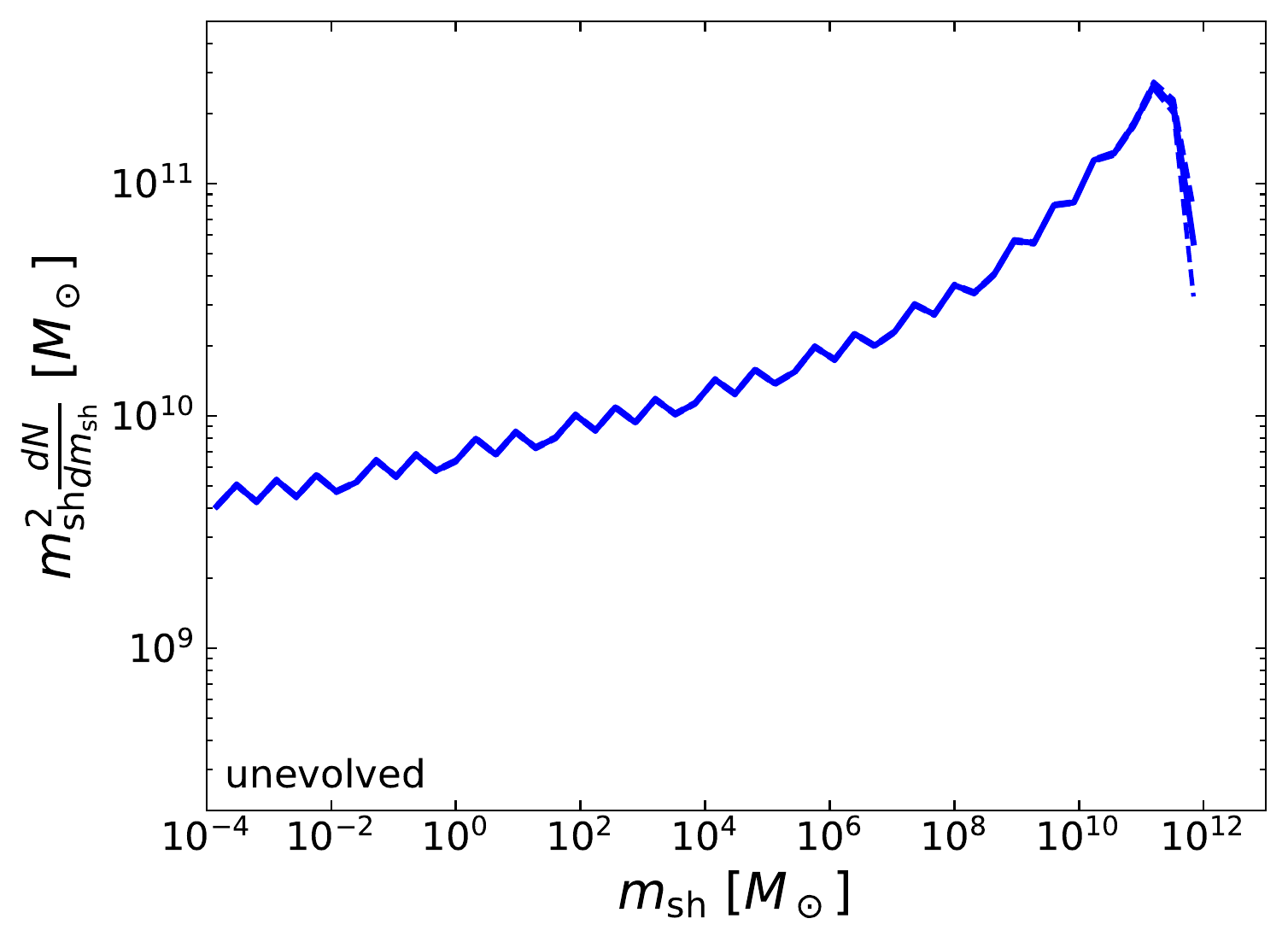}
\end{minipage}
  \begin{minipage}{0.45\hsize}
  \includegraphics[bb=7 7 430 318,scale=0.5]{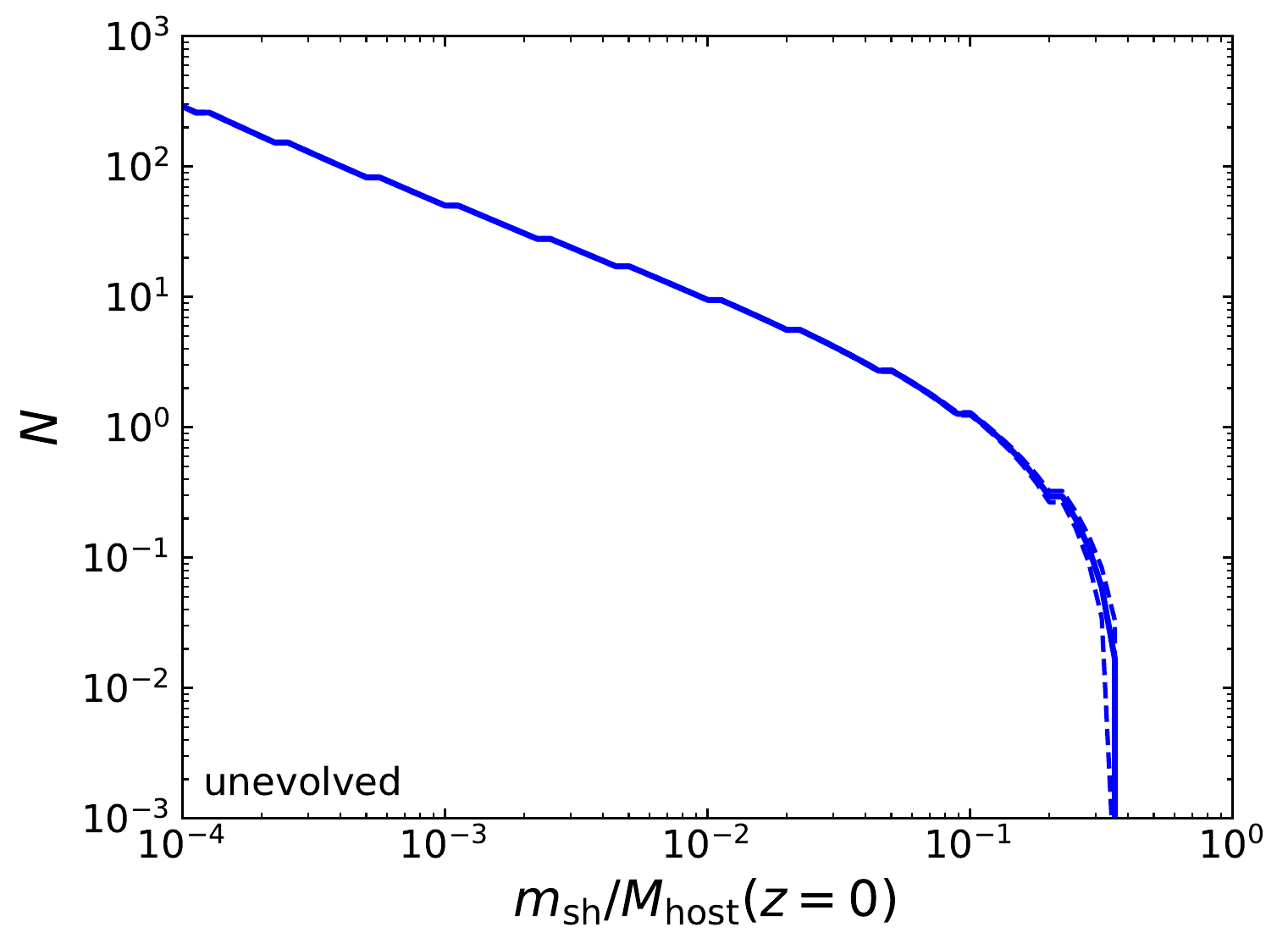}
  \end{minipage}
\end{tabular}
\caption{Differential (left) and cumulative (right) unevolved mass function at $z=0$ for host of $M(z=0)=1.30\times10^{12}M_\odot$. 
The solid (dashed) line shows the average ($1\sigma$ scatter) for 500 realizations of the host. Non-smooth features are induced by the binning effect and can be removed when we change the mass bins for plotting.}
\label{f:unevmassfcn_single}
\end{figure*}

Figure~\ref{f:unevmassfcn_single} shows the unevolved mass function at $z=0$ derived in the above procedure. In the left panel, we show the differential mass function which covers a wider range for the subhalo of  $m=[10^{-4}M_\odot,10^{12}M_\odot]$, while the cumulative mass function in a narrower range of $m/M_{\rm host}(z=0)=[10^{-4},1]$ in the right panel. We take 50 and 100 mass bins for each mass function, respectively. The wavy features in the panels are caused by the binning effect. 
The solid (dashed) line corresponds to the average ($1\sigma$ scatter) of 500 host realizations in each panel. The scatter of the mass function which originates from the different host evolution histories is small and its amount is comparable to those obtained in numerical simulations. The comparison with the numerical subhalo mass functions is shown in Sec.~\ref{ss:hostscatter}.  

\subsection{Evolved mass function}
\label{ss:ev}
Unevolved mass functions obtained in Sec.~\ref{ss:unev} is then combined with the analytical models of the tidal evolution to determine the evolved mass functions. The model of \citet{Hiroshima:2018kfv} is based on the orbit-averaged mass-loss prescriptions~\citep{Jiang:2013kza,Jiang:2014nsa} and capable of treating the dependence of the mass-loss rate on the host mass and the redshift. We starts the calculation from the data of the host evolution and the accreting mass distribution generated in the previous sections. The data is in the form of the $(N_{\rm sh},N_z)$ matrix, tracking the history from high to low redshifts in the following manner:
\begin{enumerate}

\item Convert the $M_{200}$ mass, which is adopted in previous sections, to the virial one\footnote{Functions of $\delta(z)$ and $\sigma(M)$ in \citet{Ludlow:2016ifl} are given in terms of $M_{200}$ while the tidal prescription in this section is based on the virial quantities. The transformation from $M_{200}$ to $M_{\rm vir}$ keeps the consistency of the formalism.}. The concentration-mass relation of \citet{Correa:2014xma,Correa:2015dva} are applied to determine the virial mass of host and subhalos. Assuming the NFW halo profile~\citep{Navarro:1995iw}, we use the following quantities 
\begin{eqnarray}
r_{\rm vir}&=&\left(3m_{\rm vir}\times\left(4\pi\rho_c(z)\varDelta(z)\right)^{-1}\right)^{1/3}, \\
c_{\rm vir}&=&c_{200}\frac{r_{\rm vir}}{r_{200}}, \\
r_{\rm s}&=&r_{\rm vir}/c_{\rm vir}, \\
\rho_{\rm s}&=&\frac{m_{\rm vir}}{4\pi r_{\rm s}^3 f(c_{\rm vir})},\\
r_{\rm max}&=&2.163r_{\rm s},\\
V_{\rm max}&=&\sqrt{\frac{4\pi G\rho_{\rm s}}{4.625}}r_{\rm s}.
\end{eqnarray}
The $\varDelta(z)$ which describes the overdensity corresponding to the virial mass is taken from \citet{Bryan:1997dn}. By assuming the NFW model, the halo profile is described with two parameters: the characteristic density $\rho_{\rm s}$ and the scale radius $r_{\rm s}$. The set of parameters $(\rho_{\rm s},r_{\rm s})$ has one-to-one correspondence with the set $(V_{\rm max}, r_{\rm max})$. The function $f(c)=\ln(1+c)-c/(1+c)$ relates the NFW profile parameter and the halo mass in a direct way.
\item Track the evolution of accreted halos starting from $z=10$. The part of the mass matrix of subhalo corresponding to those have ever accreted before the redshift step of the interest enters the calculation. We calculate the tidal mass loss during the interval $[z+\Delta z,z]$ as
\begin{equation}
\Delta m(z)=\frac{\Delta z}{(1+z)^2H(z)}\times\left[A(M_{\rm host},z)\frac{m}{\tau_{\rm dyn}}\left(\frac{m}{M_{\rm host}(z)}\right)^{\zeta (M_{\rm host},z)}\right].
\label{eq:tidal}
\end{equation} 
 The host mass is evaluated at each redshift of the calculation by directly referring to the host evolution data. We take the time-dependent mass-loss model of \citet{Hiroshima:2018kfv} as our fiducial one. The factor $1/(1+z)/H(z)$ appears to use the redshift as the time variable for the mass loss. We discuss the influence of the uncertainty in the tidal evolution model in Sec.~\ref{ss:tidalmodel}. 
\item Replace the subhalo mass matrix with that of after the tidal mass-loss, i.e., the matrix of $m(z)$ is replaced with $m(z)-\Delta m$. If $\Delta m(z)\geq m(z)$, we set the mass of that halo to zero.
\item Go to the next redshift step of $z-\Delta z$. The treatment is the same as that of the one-step before, but the mass matrix describing subhalos those have accreted by redshift $z$ is already replaced with those after the tidal mass-loss down to that redshift step. Also, a new part of the mass matrix corresponding to those accrete between $[z,\ z-\Delta z]$ enters the calculation. The process is repeated to $z=0$. This treatment enables us to calculate the mass-loss of subhalos including the information about their accretion redshift.
\end{enumerate}
After finishing the calculation, we obtain the mass matrix of all the subhalos which have accreted between $z=[10,0]$ and evolved. 
The $V_{\rm max}$ and $r_{\rm max}$ for NFW halo before and after the tidal stripping are related as~\citep{Penarrubia:2010jk}
\begin{eqnarray}
\frac{V_{\rm max}(z=0)}{V_{\rm max}(z_{\rm acc})}&=&\frac{2^{0.4} x^{0.3}}{(1+x)^{0.4} }, \\
\frac{r_{\rm max}(z=0)}{r_{\rm max}(z_{\rm acc})}&=&\frac{2^{-0.3} x^{0.4}}{(1+x)^{-0.3}},
\end{eqnarray}
where $x=m(z=0)/m(z_{\rm acc})$. 
The relationships for $(V_{\rm max},r_{\rm max})$ before and after the tidal stripping are recasted to those for $(\rho_{\rm s},r_{\rm s})$. Using $m$, $r_{\rm s}$, and $\rho_{\rm s}$ at $z=0$, we evaluate the truncation radius $r_t=c_tr_{\rm s}$ as 
\begin{equation}
m=4\pi\rho_{\rm s} r_t^3 f(c_t).
\end{equation}
If $c_t<0.77$, we omit those halos regarding them to be fully disrupted~\citep{Hayashi:2002qv}. The mass matrix of the evolved subhalo is integrated over the redshift with the same weight as that for unevolved mass function after this cut. 

\begin{figure*}
\begin{tabular}{cc}
\begin{minipage}{0.45\hsize}
  \includegraphics[bb=7 7 428 311,scale=0.5]{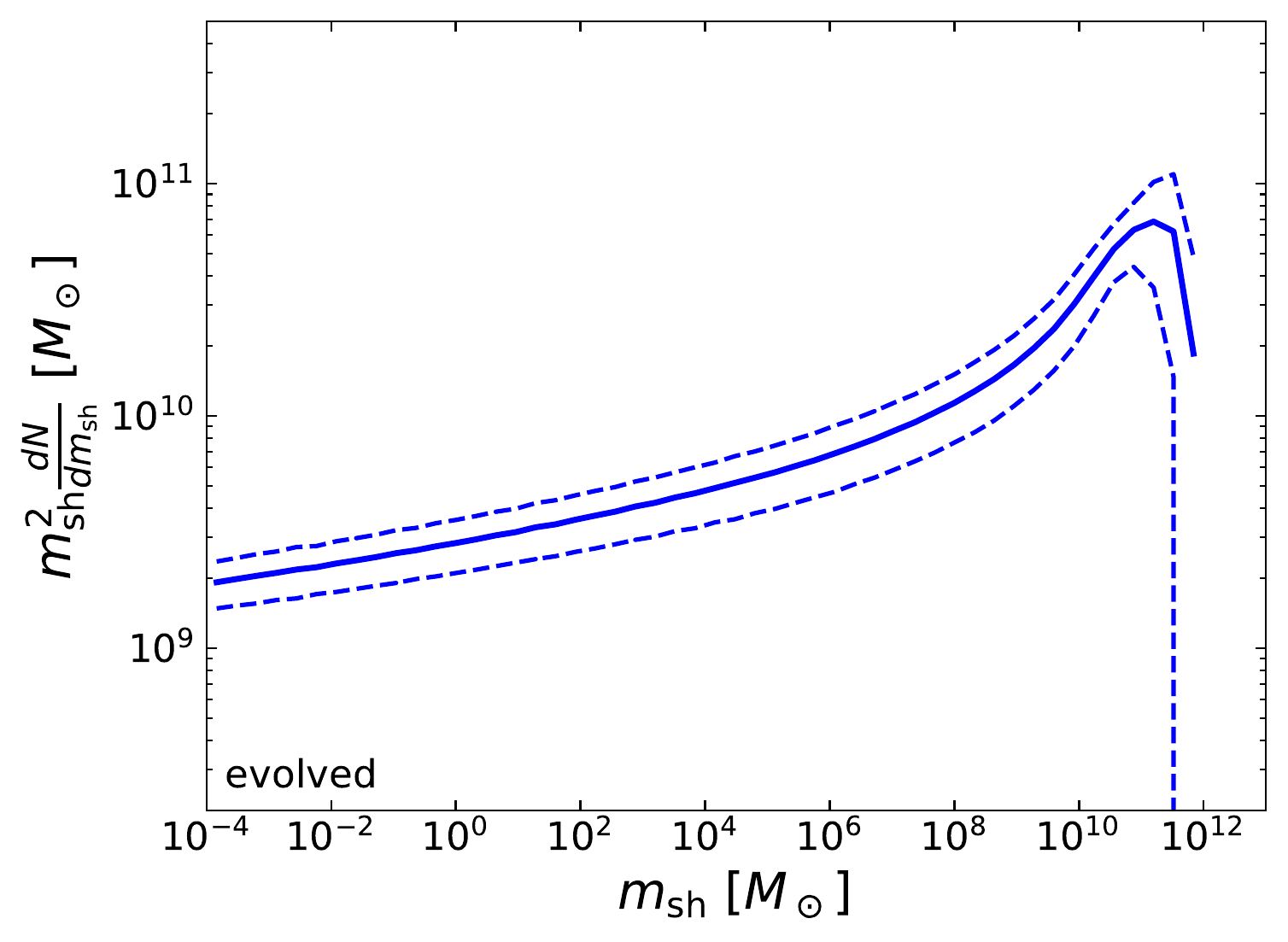}
  \end{minipage}
  \begin{minipage}{0.45\hsize}
  \includegraphics[bb=7 7 430 318,scale=0.5]{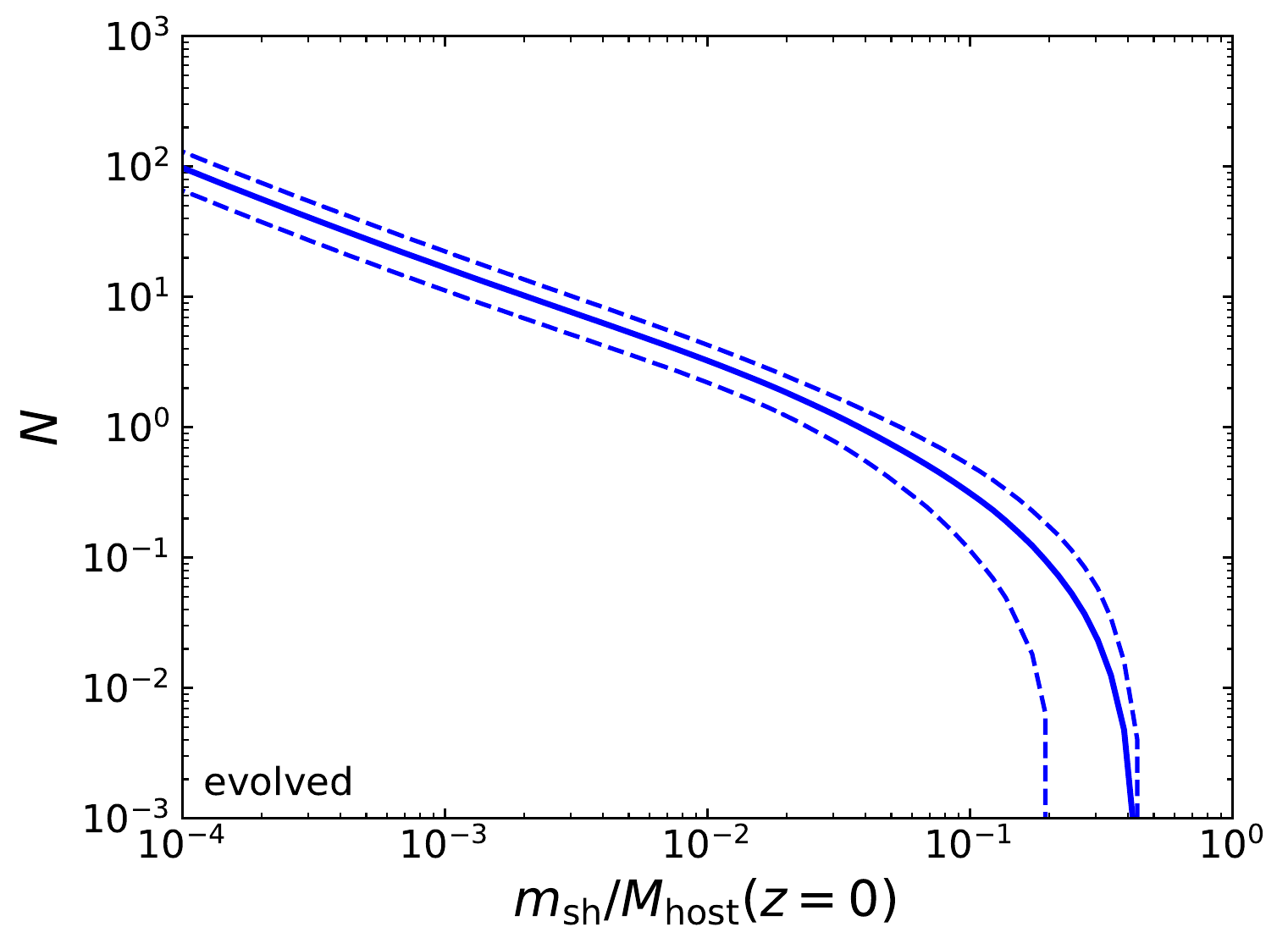}
  \end{minipage} \\
\end{tabular}  
\caption{The same as Fig.~\ref{f:unevmassfcn_single} but for evolved subhalos. The scatter becomes apparent reflecting the different host evolution histories. In this figure, the mass is measured in the virial mass for both the host and subhalos.}
\label{f:evmassfcn_single}
\end{figure*}

We show the evolved subhalo mass functions in the virial mass unit in Fig.~\ref{f:evmassfcn_single}. The host mass in the virial unit is $M_{\rm vir}(z=0)=1.5\times10^{12}M_\odot$, which is the same as that for unevolved mass function of $M_{\rm host }(z=0)=1.30\times10^{12}M_\odot$ in the $M_{200}$ unit. 
The number distribution of the accreting subhalo is the same as that for unevolved mass function. We just re-count the subhalos in each mass bin at $z=0$ using the same assignment for the mass bins with that of Fig.~\ref{f:unevmassfcn_single}. The evolved mass function is composed of all the survived subhalos which have been accreted and evolved between $z=10$ and $=0$. While we only consider the subhalos which have accreted in $z\leq10$, the extension of the calculation to a higher redshift does not change the results. This is due to the fact that subhalos that have accreted in such a high redshift are fully disrupted and do not contribute much to the mass function.
 After the tidal evolution, the total number of the subhalo decreases. Also, different from the unveolved cases, the scatter of the subhalo mass function becomes apparent, reflecting the different host evolution histories. The height of the subhalo mass function is similar to those in previous works~\citep{Jiang:2016yts,Hiroshima:2018kfv}.

\section{Quantifying the source of the scatter}
\label{s:application}
In the previous section, we have explained our basic formalism to derive the unevolved- and evolved- mass functions. Those mass functions depend only on the physical properties of DM halos, which are described in the EPS theory. However, those in either of the observations or simulations include extrinsic contributions such as the Poisson fluctuation or the host-mass scatters. We first discuss the Poisson fluctuation and the host-mass scatter effects separately to quantify each contribution. Then we show the mass function including both effects, which should be the one compared to those in observations or simulations. This could be a new scheme to extract intrinsic property from observables. The uncertainties from the treatments of the tidal disruption are also discussed in this section.

\subsection{The effect of the Poisson fluctuation}
\label{ss:poisson}
The Poisson fluctuation in the number count of the subhalo at the massive end is inevitable for both in the simulations and observations. By removing the contributions from the Poisson fluctuation, we can retrieve the nature of DM halo from observables. We provide a mapping between the subhalo mass function with and without the Poisson fluctuation by implementing this effect to our mass function derived in Sec.~\ref{s:shmf} in the following way:

\begin{enumerate}
\item Construct the cumulative subhalo function for each realization (see Sec.~\ref{s:shmf}). The cumulative mass function is normalized by dividing it by the total number of the subhalo, $N_{\rm tot}$. 
\item Generate a random number following the Poisson distribution of which expectation value is $N_{\rm tot}$. We denote this generated number as $N_{\rm tot,P}$.
\item Generate $N_{\rm tot,P}$ random numbers assuming a uniform distribution in the interval of $[0,1]$, then convert the random numbers to the halo mass using inverse transform method. This step corresponds to generate a new subhalo distribution of which total number is $N_{\rm tot, P}$ and the distribution function is the same as that determines the $N_{\rm tot}$.
\item Construct the mass function using the re-sampled $N_{\rm tot,P}$ subhalos.
\item Repeat the above procedure for $N_{\rm MC}=500$ times for each $N_{\rm tot}$ (i.e. the host realization).
\end{enumerate}

\begin{figure*}
\begin{tabular}{cc}
\begin{minipage}{0.45\hsize}
  \includegraphics[bb=7 7 430 318,scale=0.5]{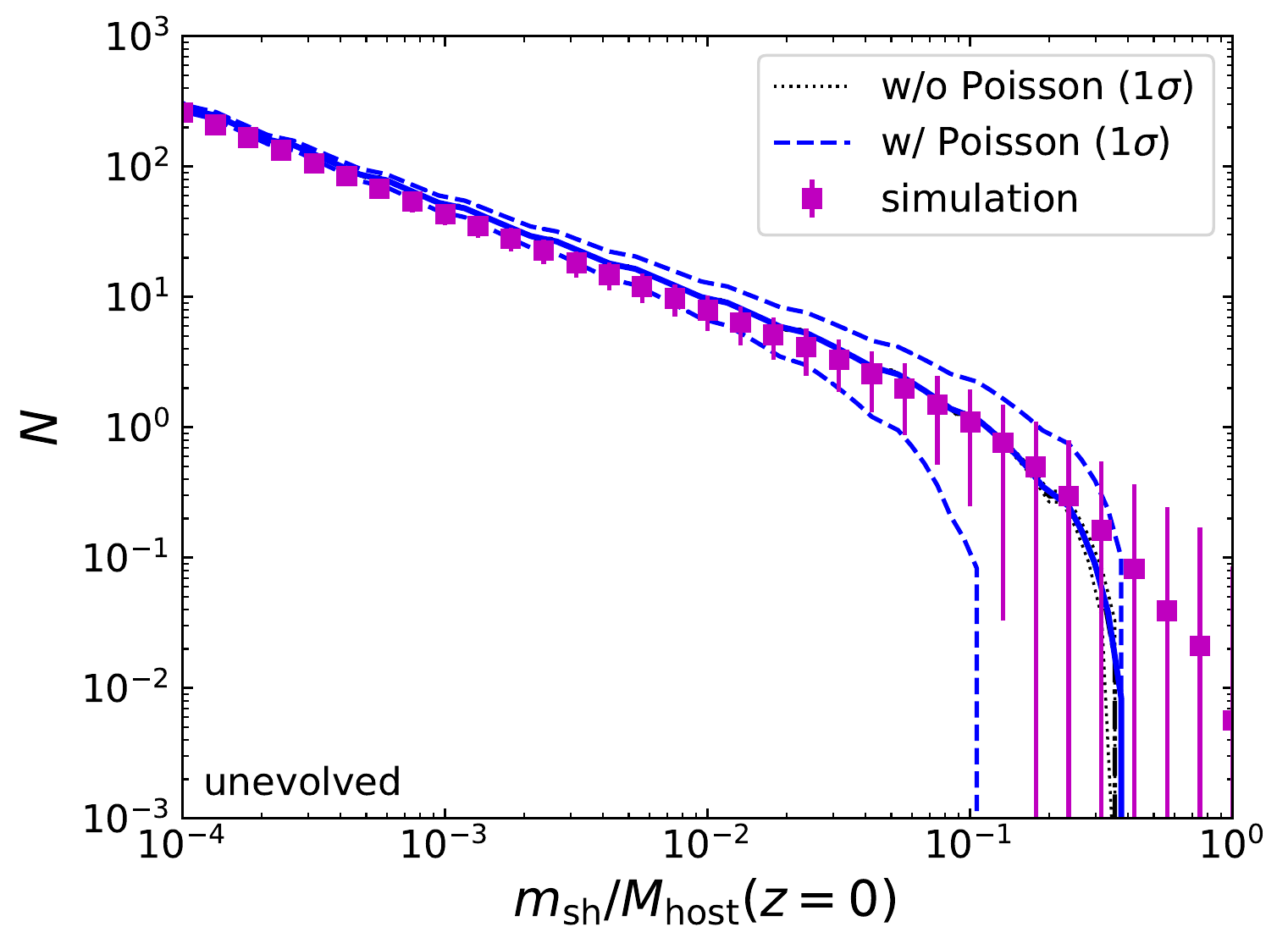}
  \end{minipage}
  \begin{minipage}{0.45\hsize}
  \includegraphics[bb=7 7 430 318,scale=0.5]{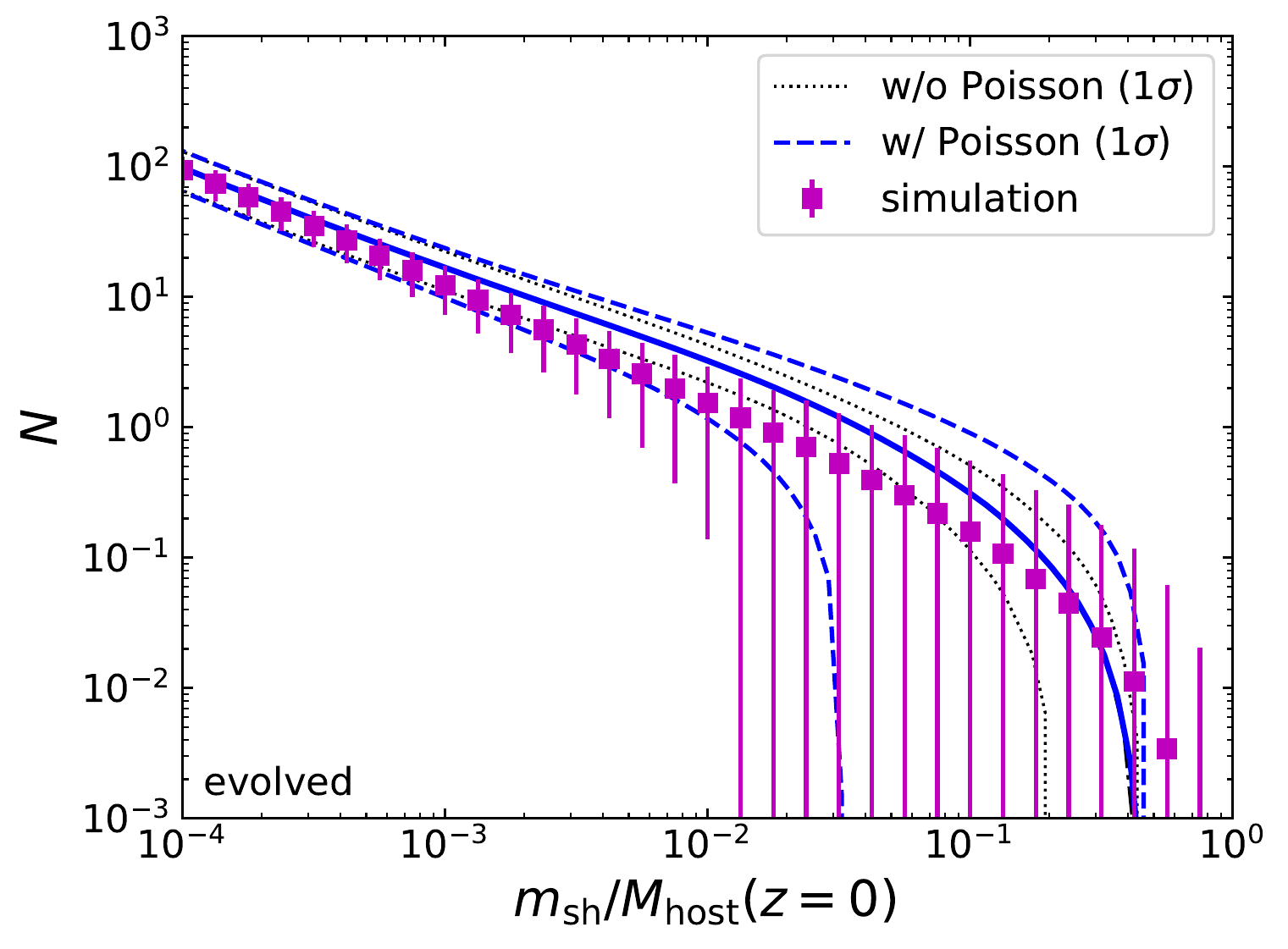}
  \end{minipage} 
\end{tabular}  
\caption{The cumulative mass function of unevolved (left) and evolved (right) subhalos at $z=0$ including the Poisson fluctuation effects. The blue solid lines show the average, while the dashed lines show the $1\sigma$ range. We also plot the mass function without the Poisson effects with black lines, of which $1\sigma$ range is visible in the dotted curve for the evolved case, while it degenerates with the average for the unevolved one. The average without the Poisson effect underlies the solid line. Square points with error bars are results from the $N$-body simulations of Sec.~\ref{s:sim}.}
\label{f:massfcn_MCsingle}
\end{figure*}

We show the cumulative mass function including the Poisson fluctuation in Fig.~\ref{f:massfcn_MCsingle}. 
The host mass at $z=0$ is the same as previous figures. 
The left and right panel corresponds to the unevolved and evolved mass function, respectively. We limit the regime of the subhalo mass in this figure to clearly see the Poisson fluctuation effects. As it is seen in Fig.~\ref{f:massfcn_MCsingle}, the $1\sigma$ scatter of the number counts becomes larger by introducing the effect. The Poisson fluctuation dominates the scatter at $N_{\rm tot}\lesssim 10$. In terms of the mass ratio, this corresponds to $m/M\gtrsim{\cal O}(10^{-2})$ for evolved subhalos, which is in a similar range pointed out in the previous work based on numerical simulations~\citep{Boylan-Kolchin:2009ztl}. 
In Fig.~\ref{f:massfcn_MCsingle}, we also plot the results from the numerical simulations of Sec.~\ref{s:sim} with square points. 
The error bar denotes the $1\sigma$ scatter. As it is seen in the figure, 
the analytical subhalo mass functions show good agreements with those in numerical simulations when the Poisson fluctuation effects are included. However, the analytical subhalo mass function shows a sharp cutoff at $m/M\sim0.5M_{\rm host}$. This feature originates from our treatment allowing only the subhalos of $m/M(z)$ at each accretion redshift. The prediction from the analytical calculation is slightly larger compared to the numerical ones. Our assumption that to determine the normalization of the accreting halo distribution would be responsible for this difference. In this picture, smooth accretion of the mass could not be included and hence the mass fraction to be assigned in subhalos should get larger.

\subsection{The effect of the host-mass scatter}
\label{ss:hostscatter}
The scatter of the host-mass estimate also contributes to the observed subhalo mass function. 
In our calculation, the host mass can be fixed to an arbitrary single value. However, simulated halos distribute in a finite mass range so careful treatments are required. The mass estimate from observations contains errors in a certain amount (e.g.~\citet{Karukes:2019jwa}) which should propagate to the subhalo mass functions. We take the contributions from the uncertainty in the host-mass estimate.

\begin{figure*}
\begin{tabular}{cc}
\begin{minipage}{0.45\hsize}
  \includegraphics[bb=7 7 430 318,scale=0.5]{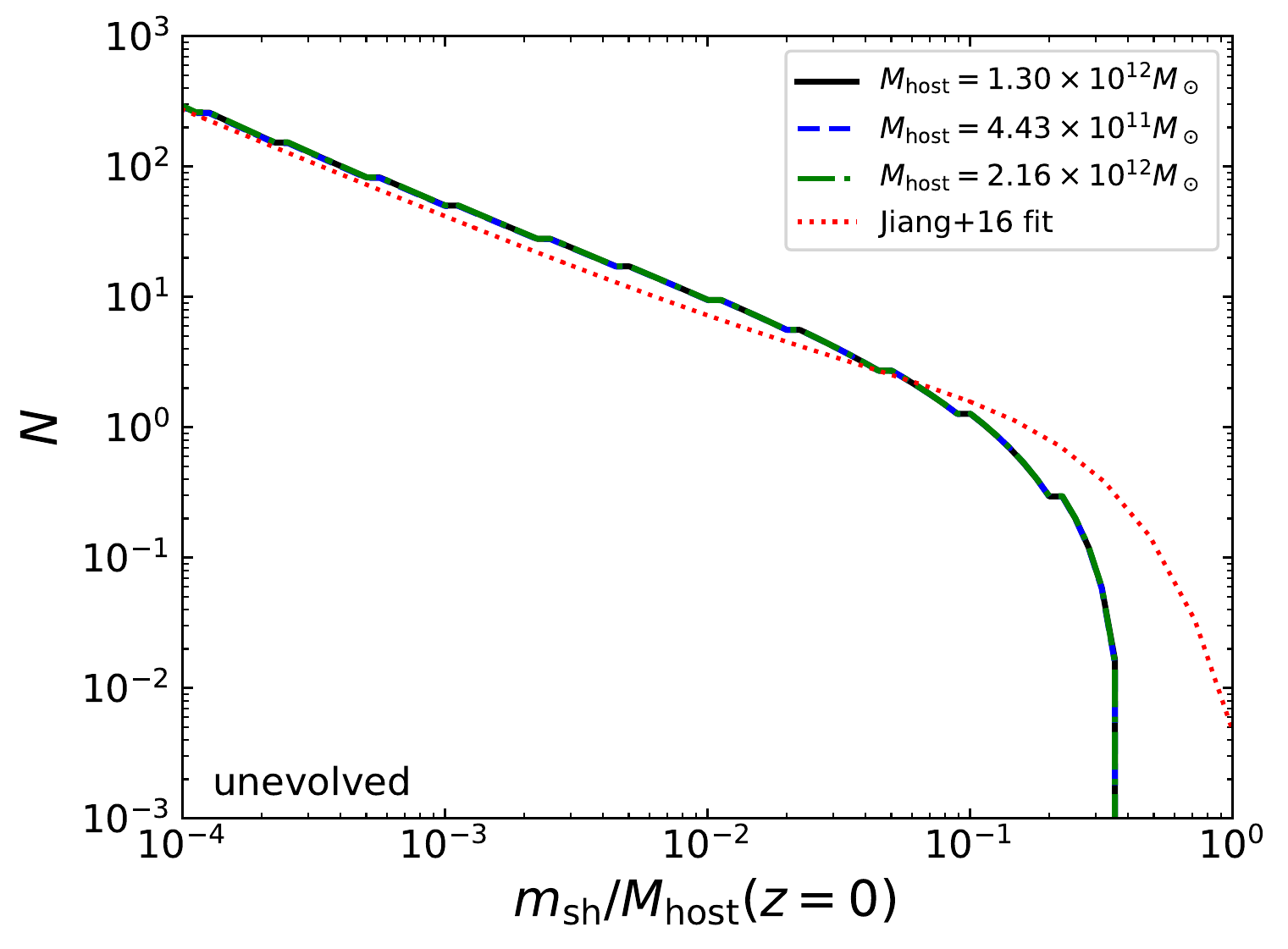}
  \end{minipage}
  \begin{minipage}{0.45\hsize}
  \includegraphics[bb=7 7 430 318,scale=0.5]{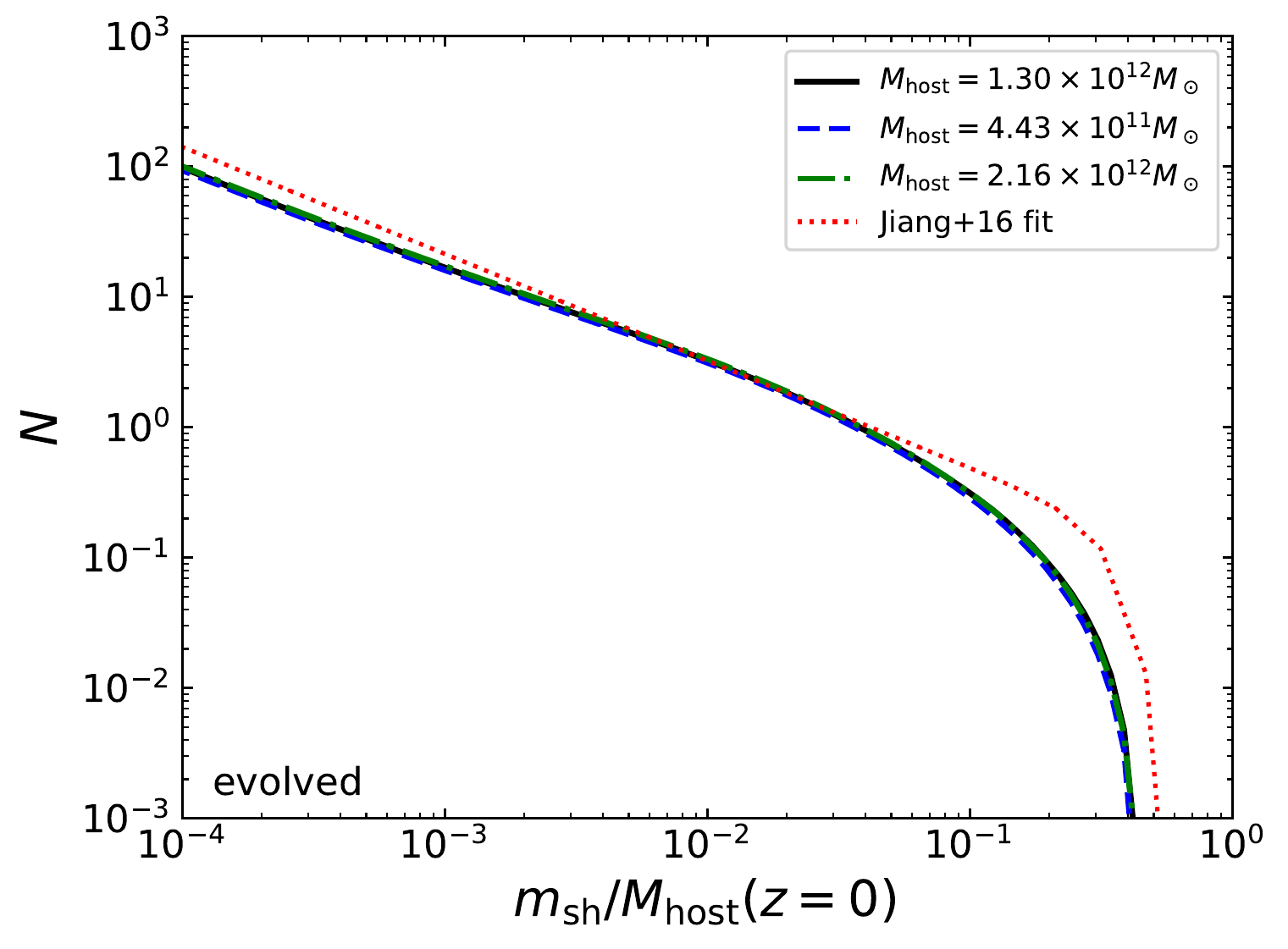}
  \end{minipage} \\
\end{tabular}  
\caption{Unevolved (left) and evolved (right) subhalo mass function for $M_{200}(z=0)=4.43\times10^{11}M_\odot$ (black, solid), $1.30\times10^{12}M_\odot$ (blue, dashed), and $2.16\times10^{12}M_\odot$ (green, dot-dashed). Each line shows the average of the 500 host realizations of that host value. The mass function with different host mass overlaps each other. The red-dotted line in each panel corresponds to the fitting formula of \citet{Jiang:2014nsa} assuming $M_{200}=1.3\times10^{12}M_\odot$. For the formula of the evolved mass function, we estimate the halo formation time as that the average host mass is the half of the current value.}
\label{f:HGcum}
\end{figure*}

In addition to our fiducial host of $M_{200}(z=0)=1.3\times10^{12}M_\odot$, we calculate the subhalo mass functions for host of $M_{\rm host}=4.43\times10^{11}M_\odot$ and $2.16\times10^{12}M_\odot$. The results are shown in Fig.~\ref{f:HGcum}. 
The left and right panels corresponds to the unevolved and evolved cases, respectively. A factor of $\sim5$ difference in the host mass estimate affects the cumulative mass function negligibly when it is plotted in terms of the mass ratio, $m/M$. The behavior is reasonable since we do not put assumptions that could significantly change the physics of the host and/or subhalo in the variation scales of the host mass. 
The results quantitatively agree with the previous works. In the figure, we plot the fitting function provided in \citet{Jiang:2014nsa} with red-dotted lines for comparison. The halo formation time which appears in the fitting function is evaluated as the redshift where the average host mass is the half of the current value. The relatively sharp cutoff compared to the fitting formula originates from our formulation that does not account for mergers of halos with comparable mass~(\citet{Moreno:2007wu} and references therin). 

\begin{figure*}
\begin{tabular}{cc}
\begin{minipage}{0.45\hsize}
  \includegraphics[bb=7 7 428 311,scale=0.5]{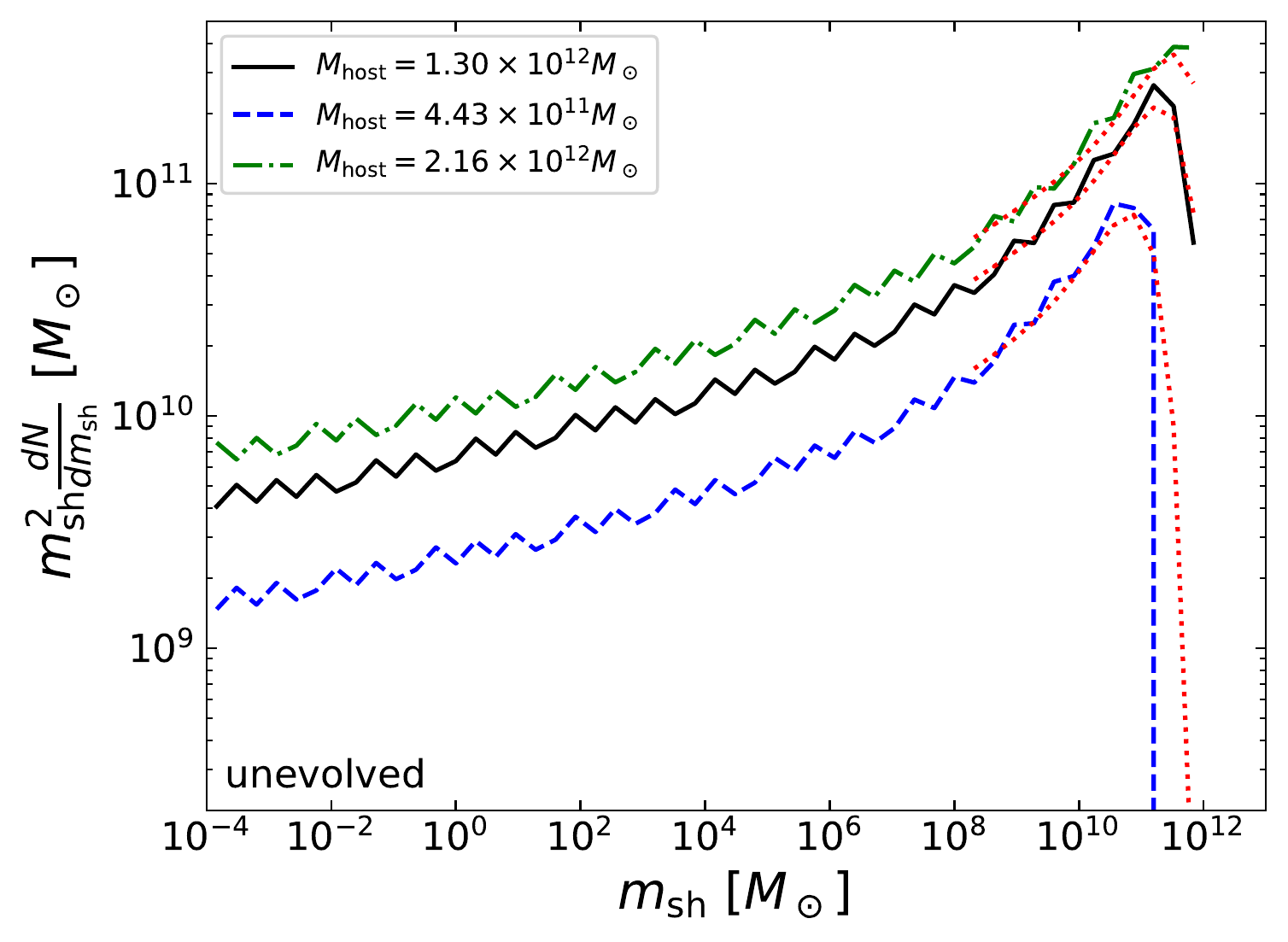}
  \end{minipage}
  \begin{minipage}{0.45\hsize}
  \includegraphics[bb=7 7 428 311,scale=0.5]{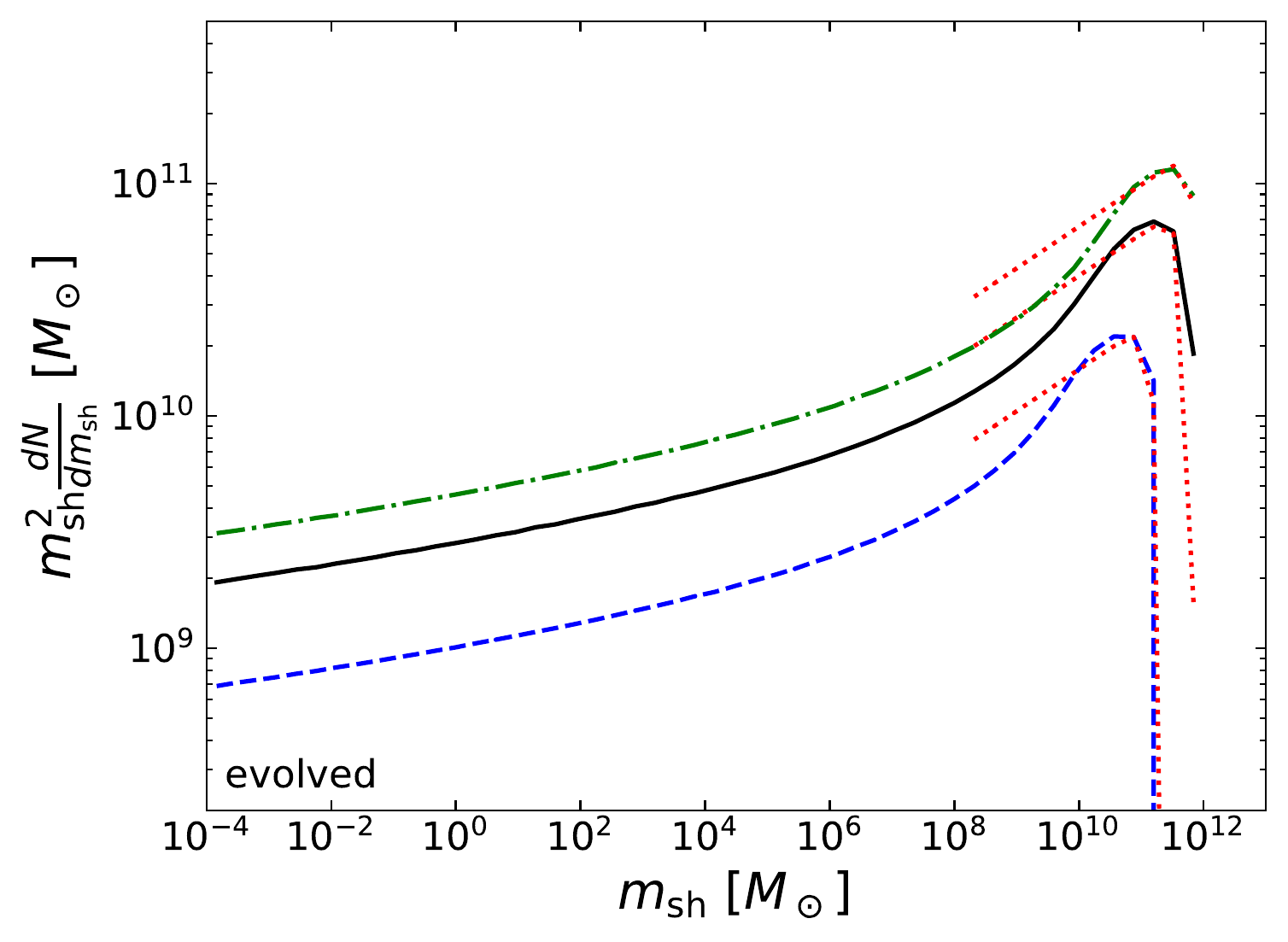}
  \end{minipage} \\
\end{tabular}  
\caption{Differential mass function of unevolved (left) and evolved (right) halos. The black solid line corresponds to the case of the $M_{\rm host}=1.30\times10^{12}M_\odot$, which is the center value we have adopted. The blue-dashed is for $M_{\rm host}=4.43\times10^{11}M_\odot$ and green dot-dashed is for $M_{\rm host}=2.16\times10^{12}M_\odot$. The peak position of the subhalo mass is determined by the host mass. The host difference between the subhalo mass function roughly scales as that of the host mass.  The red-dotted lines correspond to the formulae of \citet{Jiang:2014nsa} calculated in the same way with those of Fig.~\ref{f:HGcum}}. 
\label{f:maxmin}
\end{figure*}

We also examine how the uncertainty in the fiducial value of the host mass affect the subhalo mass function, 
by comparing the subhalo mass functions for different host masses in terms of the subhalo mass.
As it is shown in Fig.~\ref{f:maxmin}, subhalo mass function scales as the mass of the host. The host-mass dependence in obtained in our scheme is similar to that in previous works. The fitting function of \citet{Jiang:2014nsa} plotted in this figure explicitly shows the matching between the results. The maximum mass allowed for the subhalo increases for more massive hosts, hence the peak of the subhalo mass shifts with the host mass.

\subsection{Uncertainties from the tidal model}
\label{ss:tidalmodel}
In the above calculations, we have adopted the tidal evolution model for subhalo, which is characterized with two parameters: the amplitude $A$ and the dependence to the mass-ratio to the host $\zeta$ (see Eq.~\ref{eq:tidal}). The tidal evolution are still under debate~\citep{vandenBosch:2018tyt,Green:2019zkz}, while it is essential for relating the subhalo mass functions constructed in the EPS framework to those for observables. 
We quantify how the tidal model affects the predictions of the subhalo mass function at $z=0$ in this section.

\begin{figure}
\begin{center}
\includegraphics[bb=7 7 430 318,scale=0.5]{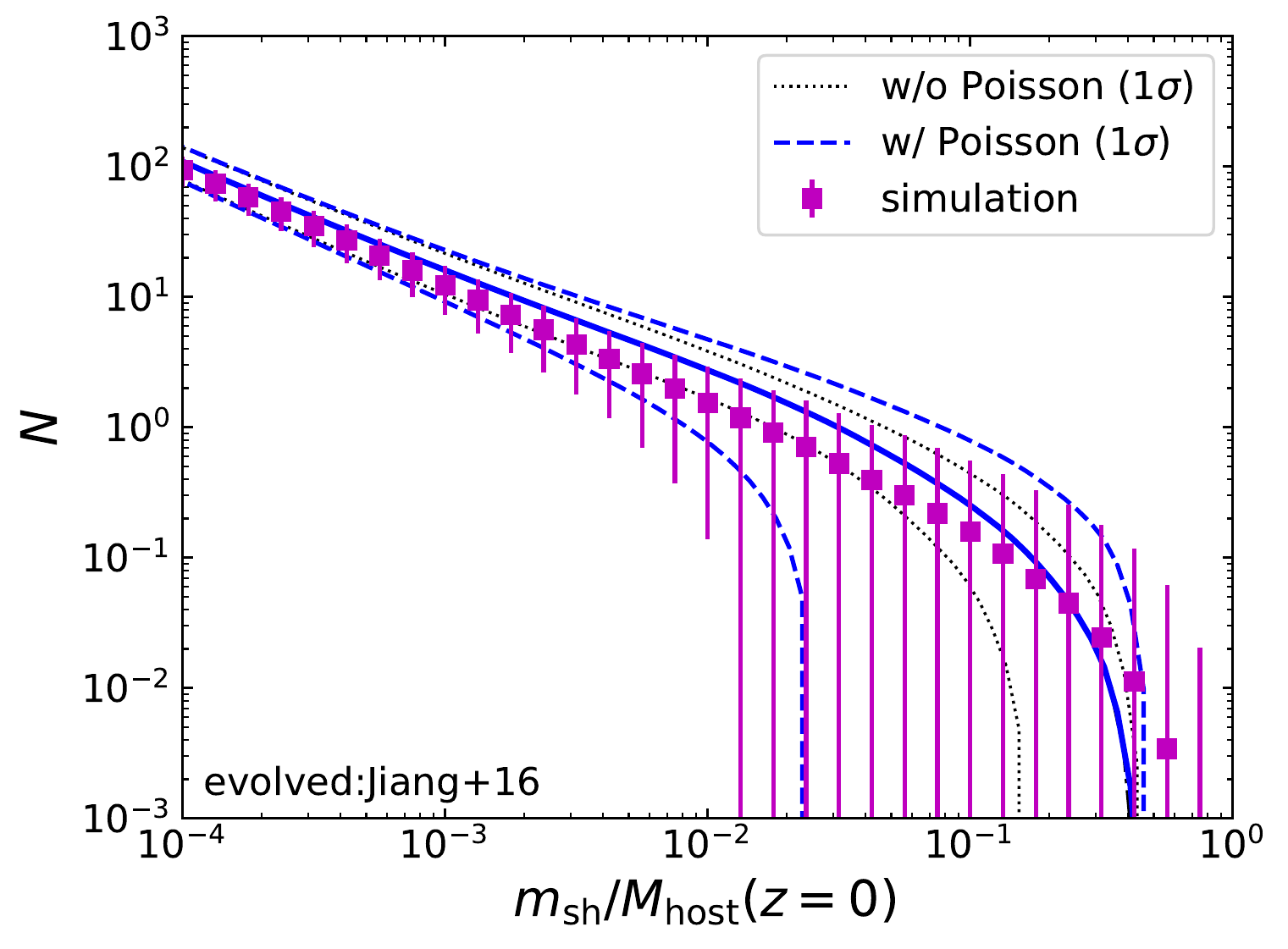}
\caption{Evolved mass function with a different model assuming time-independent parameters. 
In the range of $m/M\gtrsim{\cal O}(10^{-4})$, the difference from the result assuming time-dependent parameters is subtle. Both models show good agreement with numerical results which are shown as square with error bars.}
\label{f:vdBonly}
\end{center}
\end{figure}

For this purpose, we compare the evolved subhalo mass function assuming a time-independent parameter model of \citet{Jiang:2014nsa,Jiang:2016yts} with those predicted in the time-dependent parameter model~\citep{Hiroshima:2018kfv} which we have shown in previous sections. 
\citet{Jiang:2014nsa,Jiang:2016yts} have calibrated the model parameter $A$ and $\zeta$ with simulation~\citep{Klypin:2010qw} so that the the average and the scatter of cumulative mass function in the regime $m/M=[10^{-4},1]$ agrees, obtaining $(A,\zeta)=(0.86,0.07)$. The result is shown in Fig.~\ref{f:vdBonly}. The difference between the time-dependent and time-independent model predictions is subtle for relatively massive subhalos and both models agree well with the simulation at $m/M>{\cal O}(10^{-4})$. 

\begin{figure}
\begin{center}
\includegraphics[bb=7 7 418 315,scale=0.5]{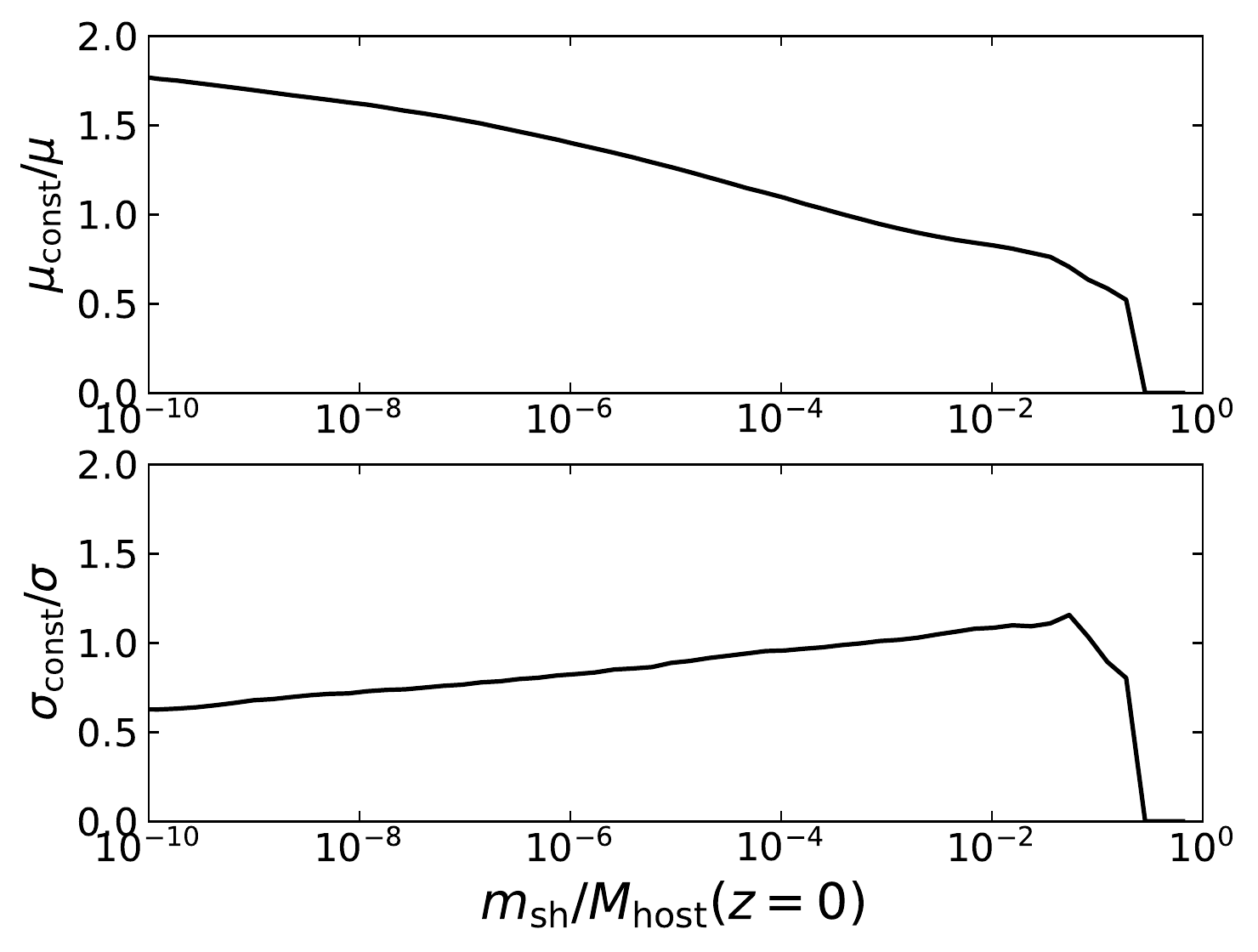}
\caption{Comparison between the cumulative mass function with two different tidal evolution models. We take the time-dependent parameter model as the reference and denoting the time-independent model with the suffix ``const" in this figure. The increasing number of the subhalo number count towards the low mass regime reflects a stronger mass-ratio dependence of the time-independent parameter model. The stronger tidal power of the low-mass host in the time-dependent parameter model generates a larger scatter in the low subhalo mass end.}
\label{f:ratioAzeta}
\end{center}
\end{figure}

The difference between two models becomes apparent when we compare in a wider mass range of 
$m/M\geq10^{-10}$, as we show in Fig.~\ref{f:ratioAzeta}. In this figure, we take   
the ratio of the mean and variance of the time-independent tidal-parameter model, which is denoted with ``const" in the figure,  
to those of the time-dependent model. A single mass host of $M_{\rm host}=1.30\times10^{12}M_\odot$ is assumed for both cases. The time-dependent parameter model predicts a smaller number of subhalos at $m/M\lesssim{\cal O}(10^{-6})$. The amount of the scatter is slightly larger for the time-independent parameter model at $m/M\gtrsim{\cal O}(10^{-3})$ while it decreases at $m/M\lesssim{\cal O}(10^{-4})$. On the other hand, the scatter of the subhalo number count remains at small mass-ratio regimes of $m/M\lesssim{\cal O}(10^{-4})$ for time-dependent parameter model. 

This reflects the fact that the trend of the model dependence is controlled by the mass-loss strength. The smaller amplitude of the tidal model parameter in this regime leads to a smaller scatter. The amplitude of the mass-loss rate is larger at $m/M\gtrsim10^{-2}$ for the time-independent parameter model hence the mean is smaller while the scatter, which is determined by the amount of the mass-loss, becomes larger. The quantitative agreement between models at $m/M\gtrsim10^{-4}$ could be an outcome for the time-independent parameter model as they have explained the tuning against the simulation in the Sec.3 of \citet{Jiang:2014nsa}.

The effects from the mass-ratio dependence, which is characterized with the parametar $\zeta$, becomes apparent in the regime of $m/M\lesssim{\cal O}(10^{-4})$ where it becomes difficult to compare with simulations. The mass-ratio dependence 
is weaker for the time-dependent parameter model, then the tidal power of the less massive hosts becomes larger compared to the case assuming the time-independent parameter. This means that the relative strength of the tidal force is larger for the former at high redshifts, i.e. in the younger hosts. Subhalos those have accreted in such a stage of the host  dominates in the low-mass regime of the mass function. 
As a result, the scatter remains at $m/M\lesssim10^{-4}$ and the prediction of the number count could be smaller by a few factors at $m/M\sim{\cal O}(10^{-10})$ for the time-dependent parameter model. 

\begin{figure}
\begin{center}
\includegraphics[bb=7 7 418 315,scale=0.5]{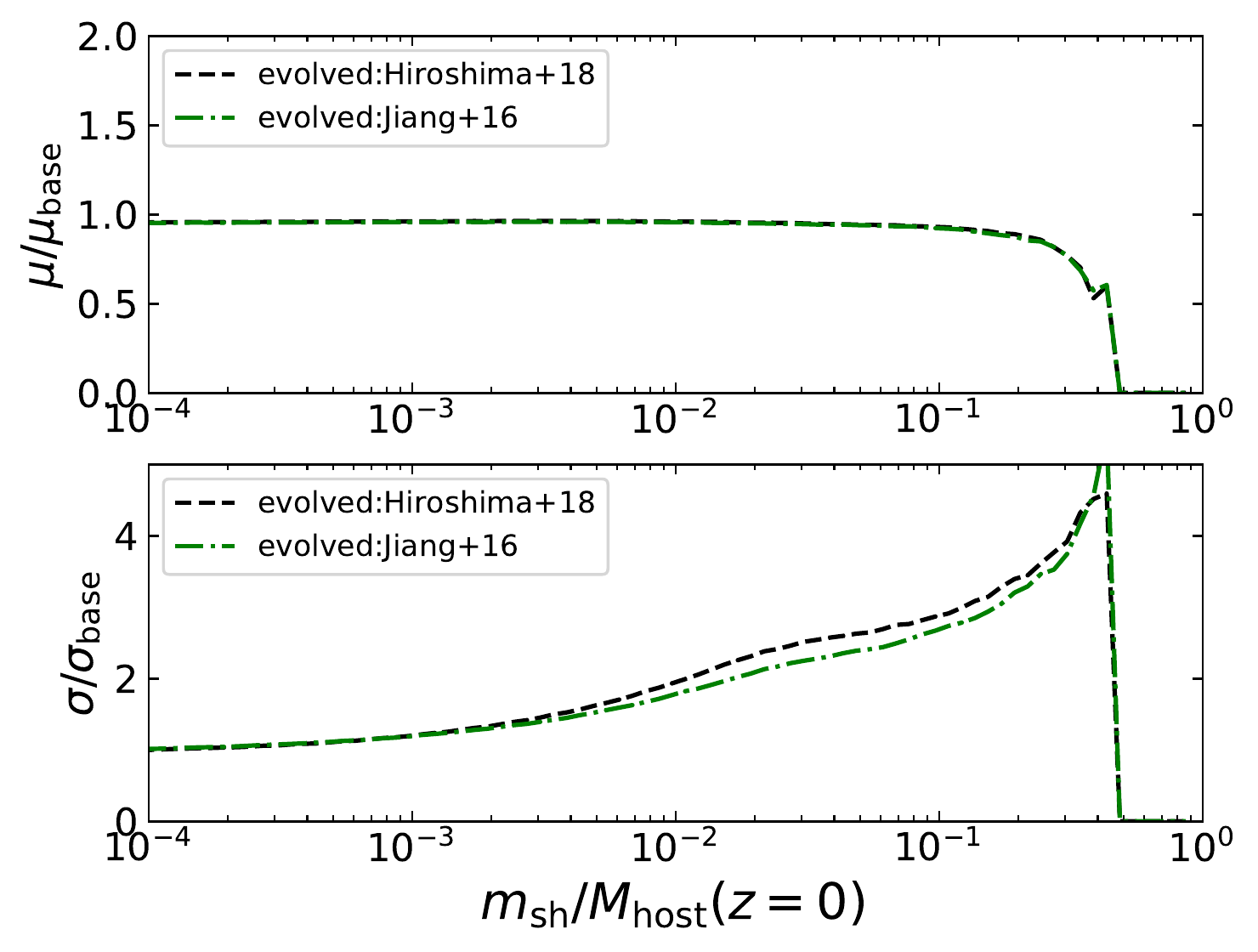}
\caption{The ratio of the mean and the variance between the predictions with- and without- the Poisson-fluctuation effects. The ``base" corresponds to the case without Poisson fluctuation. In the upper(lower) panel we show the mean (1$\sigma$ scatter). The mean is not affected in either of the tidal evolution models. 
The Poisson fluctuation affects the 1$\sigma$ scatter at $m/M\gtrsim{\cal O}(10^{-1})$. For both models, the effect is most apparent at the massive end of the subhalo mass as predicted.}
\label{f:ratioMC}
\end{center}
\end{figure}

In Fig.~\ref{f:ratioMC}, we quantify the dependence of the Poisson fluctuation on tidal models. In this figure, we take the ratio between cases with and without Poisson fluctuations. Again we consider a single mass host of $M_{\rm host}=1.30\times10^{12}M_\odot$, ignoring the host-mass scatter. The Poisson fluctuation does not affect the mean for both tidal models, while the variance at the massive end is apparently changed. The variance is enhanced by several factors at $m/M\gtrsim{\cal O}(10^{-1})$, where the regime corresponds to $N_{\rm tot}\lesssim$ a few. The amount of the Poisson fluctuation is not sensitive to the tidal-evolution models. This indicates that the estimates of the intrinsic subhalo statistics from observations or simulations with this relationship is robust against the uncertainty in the tidal evolution models. 

\section{Conclusion}
\label{s:summary}
In this article, we propose a new scheme to construct subhalo mass functions in an analytical way, which bases on the well-known EPS theory. Subhalo mass functions at arbitrary mass scales and redshifts are accessible with their statistical properties in this model. No additional parameters to be calibrated with numerical simulations are required for unvevolved mass functions while the model reconstructs the numerical results well in the regime where we can compare with. The model could be a new probe for subhalo mass functions at the smallest-scales, which are difficult to investigate in numerical simulations.

In our scheme, we track the evolution of the main branch of the host merger tree and different evolution histories of the hosts are directly reflected to their subahlo mass functions. As a demonstration, we investigate the scatter in the subhalo number counts considering the Milky-Way-like galaxies of $M_{\rm host}=1.30\times10^{12}M_\odot$ at $z=0$. Our findings are:
\begin{enumerate}
\item The Poisson fluctuation dominates at $m/M_{\rm host}\gtrsim{\cal O}(10^{-2})$. The cumulative number of subhalo exceeds one in the range of $m/M\sim[3\times10^{-2},\ 5\times10^{-2}]$ if the Poisson fluctuation effect is removed, while it extends to a much wider range of $m/M\sim[10^{-2},\ 10^{-1}]$. The amount of the enhancement in the scatter due to this effect is about a factor of a few at its peak.
\item 
The host mass variation in a finite range of several factors does not affect the cumulative subhalo mass function when it is described in terms of the mass ratio $m/M$. However, the determination of the center value of the host mass itself does affect the prediction of the halo mass function. The subhalo mass function scales as the host mass difference.
\item Comparing two models for the tidal evolution, the difference between the tidal models becomes apparent in the regime $m/M\lesssim{\cal O}(10^{-4})$ where the tuning of the tidal parameter with numerical simulations becomes difficult. The expected number count of the subhalo of $m/M\gtrsim{\cal O}(10^{-10})$ in the Milky-Way-like galaxies differs by a factor of $\sim2$ between the models we have tested in this paper. The scatter in the small mass-ratio regime becomes larger when the time-dependent tidal parameter model is adopted. This feature originates from a relatively larger tidal power of the model for the hosts in their low-mass stages.
\end{enumerate}

The above-mentioned point 3 could cause an impact on the small-scale halo search in this era of the precise astrophysics. The difference between two models becomes clearer at $m/M\lesssim10^{-6}$, where the regime gravitational lensing observations are sensitive to~\citep{Gilman:2019vca,Hsueh:2019ynk}. Also, the smallest structure of DM halo is to be probed in future pulsar timing array experiment~\citep{Ishiyama:2010es,Clark:2015sha,Kashiyama:2018gsh,Lee:2021zqw,Delos:2021rqs} hence our result could affect those predictions. 
Another impact should appear in constraining the small-scale power spectrum based on the DM annihilation signals in ultra-compact minihalos~\citep{Bringmann:2011ut,Yang:2012qi,Nakama:2017qac,Delos:2018ueo,Lee:2020wfn}. The dependence on the models of the tidal evolution shifts the such constraints by factors. The precise models for tidal evolution are needed to obtain much robust picture of the small-scale structure of our Universe.

\section*{acknowledgement}
This work is supported by  JSPS
KAKENHI Grants No. JP19K23446 (N. H.), MEXT KAKENHI Grants No. JP20H05852 (N. H.), JP20H05850 (S. A.), JP20H05861 (S. A.), JP20H05245 (T. I.), and JP21H01122 (T. I.).
T. I. has been supported by MEXT as ``Program for
Promoting Researches on the Supercomputer Fugaku'' (JPMXP1020200109), JICFuS, and IAAR Research Support Program, Chiba University, Japan. 

We thank Instituto de Astrof\'isica de Andaluc\'ia (IAA-CSIC), Centro de Supercomputaci\'on de Galicia (CESGA) and the Spanish academic and research network (RedIRIS) in Spain for hosting Uchuu Data Release one in the Skies \& Universes site for cosmological simulations. The Uchuu simulations were carried out on Aterui II supercomputer at Center for Computational Astrophysics, CfCA, of National Astronomical Observatory of Japan, and the K computer at the RIKEN Advanced Institute for Computational Science. 
The numerical analysis were partially carried out on XC40 at the Yukawa Institute Computer Facility in Kyoto University.
The Uchuu DR1 effort has made use of the skun@IAA\_RedIRIS and skun6@IAA computer facilities managed by the IAA-CSIC in Spain (MICINN EU-Feder grant EQC2018-004366-P).

\section*{Data Availability}
The calculation code underlying this article are available in \url{https://github.com/shinichiroando/sashimi-c} at \url{https://doi.org/10.3390/galaxies7030068} and \url{https://doi.org/10.1103/PhysRevD.97.123002}. The data are available in the article and in \url{http://skiesanduniverses.org/} at  \url{https://doi.org/10.1093/mnras/stab1755}. 
 

\bibliographystyle{mnras}
\bibliography{refs, refs_TI}

\bsp	
\label{lastpage}
\end{document}